\documentclass[oribibl]{llncs}

\usepackage[english]{babel}
\usepackage[utf8x]{inputenc}
\usepackage{amsmath}
\usepackage{graphicx}
\usepackage{url}
\usepackage{epstopdf}
\usepackage[colorinlistoftodos]{todonotes}

\begin{document}

\mainmatter
\title{Taming the zoo - about algorithms implementation in the ecosystem of Apache Hadoop}

\author{Piotr Jan Dendek \and Artur Czeczko \and Mateusz Fedoryszak \and Adam Kawa \and Piotr Wendykier \and Łukasz Bolikowski
}
\authorrunning{Piotr Jan Dendek et al.} % abbreviated author list (for running head)
\institute{Interdisciplinary Centre for Mathematical and Computational Modelling,\\
University of Warsaw\\
\email{\{p.dendek, a.czeczko, m.fedoryszak, p.wendykier, l.bolikowski\}@icm.edu.pl, \\
kawa.adam@gmail.com}}

\maketitle  

\begin{abstract}
Content Analysis System (CoAnSys) is a research framework for mining scientific publications using Apache Hadoop. This article describes the algorithms currently implemented in CoAnSys including classification, categorization and citation matching of scientific publications. The size of the input data classifies these algorithms in the range of big data problems, which can be efficiently solved on Hadoop clusters.
\end{abstract}


\section{Introduction}
\label{sec:introduction}

Growing amount of data is one of the biggest challenges both in commercial and scientific applications \cite{McKinsey}. General intuition is that well embraced information may give additional insight into phenomena occurring in the data. To meet this expectation, Google proposed the MapReduce paradigm, which open-source implementation is Apache Hadoop. The ecosystem of Apache Hadoop gives a way to efficiently use hardware resources and conveniently describe data manipulations.
In the Centre for Open Science (CeON) we employed that solution and produced Content Analysis System (CoAnSys) – the framework for finer scientific publication mining. CoAnSys enables data engineers to easily implement any data mining algorithm and chain data transformations into workflows. During the development of CoAnSys, the set of good implementation practices and techniques has clarified \cite{MentalEq}.
In this article we share a practical knowledge in the ground of big data implementations, based on the three use cases: citation matching, document similarity and document classification.

The rest of this paper is organized as follows. Section 2 presents an overview of CoAnSys. Section 3 describes algorithms developed at CeON, which are well suited for MapReduce paradigm. Section 4 contains conclusions and future plans.

\section{CoAnSys}
\label{sec:conasys}

\begin{figure}
\begin{center}
\includegraphics[width=\linewidth]{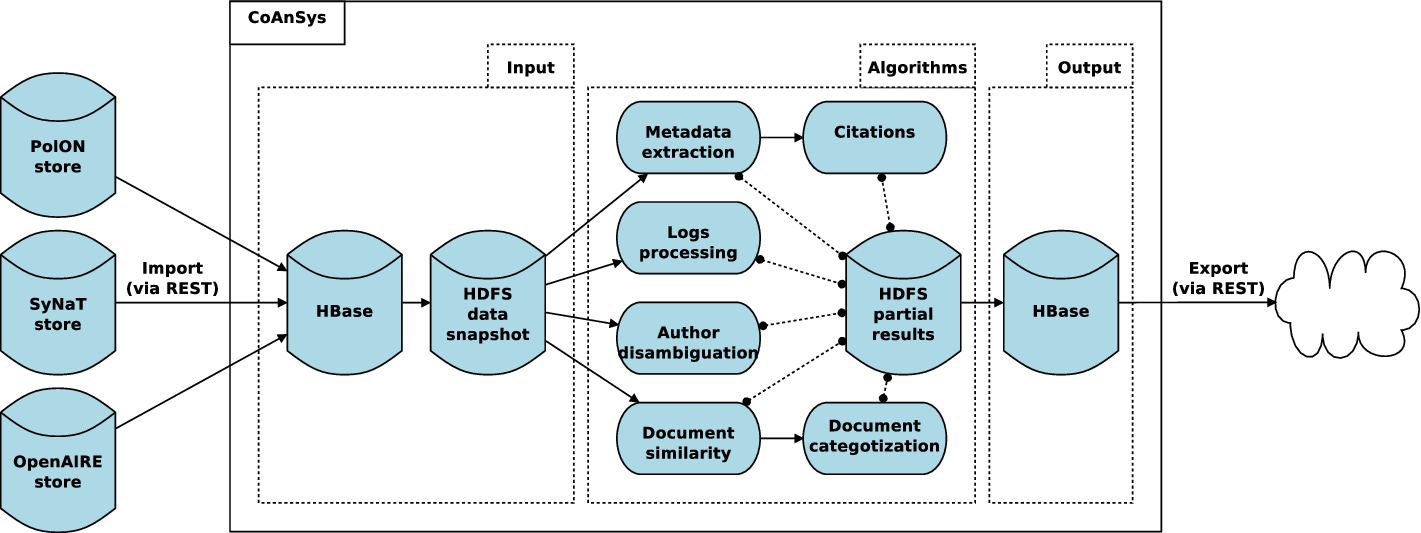}
 \end{center} 
\caption{A generic architecture of CoAnSys.}
\label{generic_architecture_coansys}
\end{figure}

The main goal of CoAnSys is to provide a framework for processing a large amount of text data. Currently implemented algorithms allow for knowledge extraction from scientific publications. Similar software systems include Behemoth \footnote{\url{https://github.com/DigitalPebble/behemoth}}, UIMA \cite{Ferrucci2004}, Synat \cite{Bembenik:1503879}, OpenAIRE \cite{Manghi2010} and currently developed OpenAIREplus \cite{Manghi2012}. The difference between CoAnSys and aforementioned tools lies in the implementation of algorithms. CoAnSys is used to conduct a research in text mining and machine learning, all methods implemented in that framework have been already published or will be published in a future. An architecture overview of CoAnSys is illustrated in Fig.\ref{generic_architecture_coansys}. 

While designing the framework, we paid a close attention to the input/output interfaces. For this purpose CoAnSys employs Protocol Buffers\footnote{\url{http://code.google.com/p/protobuf/}} - a widely used method of serializing data into a compact binary format. Serialized data is then imported into the HBase using REST protocol. This allows for simultaneous import of data from multiple clients. On the other hand, querying a large number of records from HBase is slower than performing the same operation on a sequence file stored in the HDFS. Therefore, in the input phase of CoAnSys workflow, the data is copied from an HBase table to an HDFS sequence file and such format is recognized as a valid input for the algorithms. 

Six modules currently implemented in CoAnSys are illustrated in the Algorithms box in Fig.\ref{generic_architecture_coansys}. Each module performs a series of MapReduce jobs that are implemented in Java, Pig\footnote{\url{http://pig.apache.org/}} or Scala. Apache Oozie\footnote{\url{http://oozie.apache.org/}} is used as a workflow scheduler system that chains modules together. Each module has well defined I/O interfaces in the form of Protocol Buffers schemas. This means, that sequence files are also used as a communication layer between modules. The output data from each workflow is first stored as an HDFS partial result (sequence file containing records serialized with Protocol Buffers) and then it is exported to the output HBase table where it can be accessed via REST.
 
Even though CoAnSys is still in an active development stage, there are at least three ongoing projects that will utilize parts of CoAnSys framework. POL-on\footnote{\url{http://polon.nauka.gov.pl}} is an information system about higher education in Poland. SYNAT\footnote{\url{http://www.synat.pl/}} is a Polish national strategic research program to build an interdisciplinary system for interactive scientific information. OpenAIREplus\footnote{\url{http://www.openaire.eu/}} is the European open access data infrastructure for scholarly and scientific communication.

\section{Well-suited Algorithms}
\label{sec:well_suited_algorithms}

In this section a few examples of MapReduce friendly algorithms are presented. MapReduce paradigm put a certain set of constraints, which are not acceptable for all algorithms. From the very beginning the main afford in CoAnSys have been put on document analysis algorithms, i.e. author name disambiguation \cite{DendekBL:2012:IAPR,Dendek2013,BolikowskiD:2011}, metadata extraction \cite{Tkaczyk2012}, document similarity and classification calculations \cite{Lukasik2013, Kusmierczyk2012}, citation matching \cite{Fedoryszak2013, Matfed:2013}, etc. Some of algorithms can be used in Hadoop environment out-of-box, some need further amendments and some are entirely not applicable \cite{Lin2012}.

For the sake of clarity, the description of the algorithms focuses on the implementation techniques (such as performance improvements), while the enhancements intended to elevate accuracy and precision are omitted.

\subsection{Citation Matching - General Description}
\label{sec:citation_matching_general_description}
\begin{figure}
\begin{center}
\includegraphics[width=\linewidth]{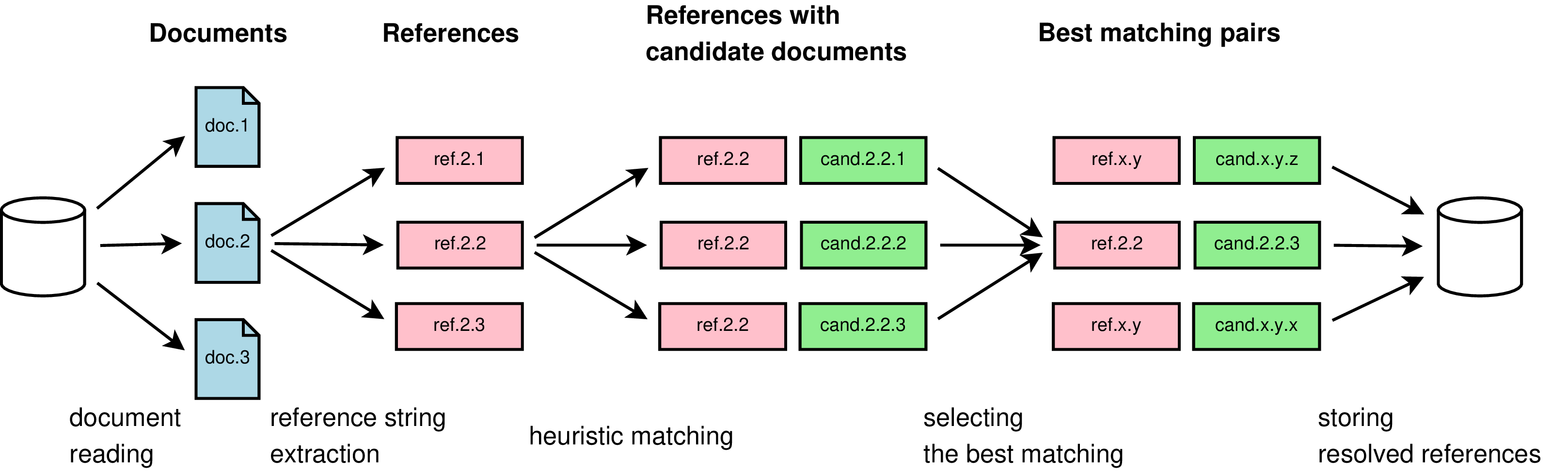}
\end{center} 
\caption{\textbf{Citation matching steps.} At first, the documents from appropriate SequenceFile are read and their metadata is extracted. Then, in the first step of citation matching, a heuristic is used to find documents that may match each citation. In the next step, the best match for each citation is selected. Finally, the results are persisted in a SequenceFile. Note that steps that transform one entry into many can be implemented as mapping and those transforming many into one as reduction.}
\label{fig:cit-match-main}
\end{figure}

A task almost always performed when researching scientific publications is the citation resolution. It aims for matching citation strings against the documents they reference. As it consists of many similar and independent subtasks, it can greatly benefit from the use of MapReduce paradigm. We may describe citation matching in the following, illustrated in Fig.\ref{fig:cit-match-main}, steps:

\begin{enumerate}
\item Retrieve documents from the store.
\item Map each document into its references (i.e. extract reference strings).
\item Map each reference to the corresponding document (i.e. the actual matching).
\begin{enumerate}
    \item Heuristically select best matching documents (map step).
    \item Among them return the one with the biggest similarity (but not smaller than a given threshold) to the reference (reduce step).
  \end{enumerate}
\item Persist the results.
\end{enumerate}

\subsection{Citation matching - Implementation Details}
\label{sec:citation_matching_implementation_details}

\subsubsection{Index}
\label{sec:index}

Heuristic matching is done using an approximate author index allowing retrieval of elements with edit distance \cite{ElmagarmidIV:2007} lesser or equal to 1. We have managed to design it to fit MapReduce paradigm and Hadoop environment in particular. It is implementing the ideas presented by Manning et al.\ in Chapter 3 of \cite{Manning2008}.

To store the index, we needed a data structure that would enable fast retrieval as well as scanning of sorted entities. Hadoop MapFile turned out to be a good solution. It extends capabilities of a SequenceFile (which is a basic way of storing key-value pairs in the HDFS) by adding an index to the data stored in it. The elements of a MapFile can be retrieved quickly, as the index is usually small enough to fit in a memory. The data is required to be sorted which makes changing a MapFile laborious, yet it is not a problem since our indices are created from scratch for each algorithm execution. Hadoop exposes an API for MapFile manipulation which provides operations such as sorting, entity retrieval and data scanning.

\subsubsection{Distributed Cache}
\label{sec:distributed_cache}

Every worker node needs to access the index (the whole index, not just a part of MapFile). It seemed, therefore, to be a good idea to store it in the HDFS. Unfortunately, this approach has a serious performance issues, because the index is queried very often. The speed of a network connection is a main bottleneck here. While seeking for a better solution, we have noticed the Hadoop Distributed Cache. It allows to distribute some data among worker nodes so that it can be accessed locally. The achieved performance boost was enormous - citation matching on the sample of 2000 documents worked four times faster.

\subsubsection{Scala and Scoobi}
\label{sec:scala_and_scoobi}

As MapReduce originates in a functional programming paradigm, one might suppose it would fit well Scala language. Indeed, during citation matching implementation, 
 we have exercised Scoobi\footnote{\url{http://nicta.github.com/scoobi/}} library which enables easy Hadoop programming in Scala by providing an API similar to Scala’s native collections. This way a very clean code can be written. In spite of great reduction of the boilerplate, Scoobi does not restrict access to some low level Hadoop features. When one desires complete control over job execution, though, the default Hadoop API may need to be used.

\subsubsection{Task Merging}
\label{sec:task_merging}
Fig.\ref{fig:cit-match-main} shows subsequent map steps (which could be implemented as MapReduce jobs with zero-reducers). Sometimes it might be beneficial to merge such tasks, as effectiveness may be improved by avoiding intermediate data storage and additional initialization cost. That is what Scoobi tends to do when computing an execution plan. This leads to a parallelism reduction,  which, in turn, can negatively impact the performance.

For instance, suppose we want to process two documents, first containing one citation and second containing fifty. In addition, let's assume that we are using the cluster of two nodes. If citation extraction and heuristic matching steps are merged, then the first mapper would extract and match one citation and the second would have to process fifty of them. On the other hand, if the tasks remain independent after citation extraction and before actual matching, a load balancing will occur. As a result, a citation matching workload will be more equally distributed.
Unfortunately, Scoobi does not allow for task merging prevention and eventually this part of the process has been implemented using the low-level Hadoop API.

\subsection{Document Similarity - General Description}
\label{sec:document_similarity_general_description}

\begin{figure}
\begin{center}
\includegraphics[width=\linewidth]{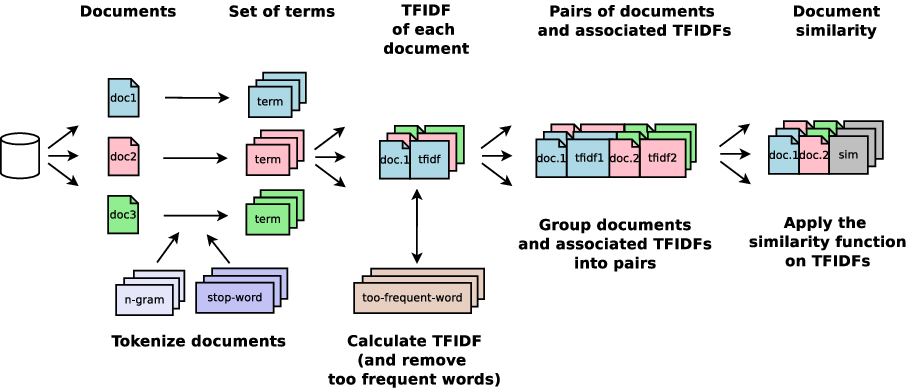}
 \end{center} 
\caption{\textbf{Document similarity steps. At first, each document is split into terms and the importance of each term to a document is calculated using TFIDF measure (resulting in the vector of weights of terms for each document). Then, documents are grouped together into pairs, and for each pair, the similarity is calculated based on the vectors of weights of common terms associated with the documents.}}
\label{document_similarity_coansys}
\end{figure}

A good illustration of the well suited MapReduce problem 
is the computation of a document similarity in a large 
collection of documents, assuming that the similarity between two documents is expressed as the similarity between weights of their common terms. Such an approach divides the computation into two consecutive steps: 
\begin{enumerate}
\item the calculation of weights of terms for each document 
\item the invocation of a given similarity function on weights of terms related to each pair of documents
\end{enumerate}
In our current implementation, the term frequency inverse-document frequency (TFIDF) measure and the cosine similarity have been used to produce weights for terms and calculate their similarity respectively. The process is briefly depicted in Fig.\ref{document_similarity_coansys}.

\subsubsection{Term Weighting Algorithm}
\label{sec:term_weight_algorithm}
TFIDF is a well-known information retrieval algorithm that measures how important a word is to a document in a collection of documents. Generally speaking, the word becomes more important to a document, if it appears frequently in this document, but rarely in other documents. 

Formally, TFIDF can be described by Eq.\ref{eq:tfidf}-\ref{eq:idf}.
\begin{eqnarray}
tfidf_{i,j} & = & tf_{i,j}*idf_{i} \label{eq:tfidf}\\
tf_{i,j} & = & \frac{n_{i,j}}{\sum_k n_{k,j}}  \label{eq:tf}\\
idf_{i} & = & log\frac{|D|}{|{d:t_i \in d}|}  \label{eq:idf}
\end{eqnarray}
where 
\begin{itemize}
\item $n_{i,j}$ is an occurrence of a term $t_i$ in document $d_j$
\item $D$ is a corpus of documents
\item $d:t_i \in d$, a document $d$ containing a term $t_i$
\end{itemize}
The way to use TFIDF in the document similarity calculation is presented in Eq.\ref{eq:cosine_doc_sim}
\begin{eqnarray}
cosineDocSim(d_x,d_y) = \frac{\sum\limits_{t_i \in d_x \cap d_y}tfidf_{i,x}*tfidf_{i,y}}{\sqrt{\sum\limits_{t_i \in d_x} tfidf^2_{i,x}} * \sqrt{\sum\limits_{t_j \in d_y} tfidf^2_{j,y}}} \;\;\;\;, for\; x < y < |D| \;\;\;\; \label{eq:cosine_doc_sim}
\end{eqnarray}

For each document $d$, this algorithm produces the vector $W_d$ of term weights $w_{t,d}$ which indicates the importance of each term $t$ to the document.
Since it consists of separate aggregation and multiplication steps, 
it can be nicely expressed in the MapReduce model with several map and reduce phases
\cite{MRAlgsCloudera, LeeHerKim:2011, WanYuXu:2009}. 
In addition, several popular techniques that increase an efficiency and a performance of the algorithm have been deployed:
\begin{enumerate}
\item stop words filtering (based on a predefined stop-word list)
\item stemming (the Porter stemming algorithm \cite{Porter:1997})
\item applying n-grams to extract phrases (considering a statistically frequent n-gram as a phrase)
\item removal of the terms with the highest frequencies in a corpus (automatically, but with a parametrized threshold)
\item weights tuning (based on the sections where a given term appears).
\end{enumerate}

\subsubsection{Similarity Function}
\label{sec:similarity_function}
Having each document represented by the vector $W_d$ of term weights 
$w_{t,d}$, one can use many well known functions (e.g inner 
product, cosine similarity) to measure similarity between 
the pair of vectors. In our implementation, we follow ideas from 
\cite{Elsayed:2008:PDS:1557690.1557767}, but provide more generic 
mechanism to deploy any similarity function that implements 
one-method interface specified by us i.e. $similarity(id(d_i), id(d_j), sort({w_{t_i,d_i}}), sort({w_{t_j,d_j}}))$ (where $id(d_i)$ denotes document id, and $sort({w_{t_i,d_i}})$ is a list of weights of common terms ordered by terms lexicographically). The similarity function receives only terms that have non-zero weights in both vectors, thus, the final score is calculated faster. This assumption remains valid only for the pairs of documents that have at least one common term.

\subsection{Document Similarity - Implementation Details}
\label{document_similarity_implementation_details}

\subsubsection{Language Choice}
Although both algorithms, TFIDF and vector similarity, can be easily expressed in the MapReduce model, they require multiple map and reduce passes, what contributes to a verbose code. In order to make the code easier to maintain and less time-consuming to implement, CeON team uses Apache Pig (enhanced by UDFs, User Defined Functions written in Java).

Apache Pig provides a high-level language (called PigLatin) for expressing data analysis programs. PigLatin supports many traditional data operations (e.g. group by, join, sort, filter, union, distinct). These operations are highly beneficial in multiple places such as stop words filtering, self-joining TFIDF’s output relations or grouping relations with a given condition.

\subsubsection{Input Dataset}
Document similarity module takes advantage of rich metadata information associated with each document. Keywords needed to compute the similarity are extracted from the title, the abstract and the content of a publication and then, they are combined with the keyword list stored in the metadata. The information in which sections a given keyword appears is taken into account during the computation of the final  weights $w_{t,d}$ in the TFIDF algorithm. A user may configure how important a given section is in the final score.

\subsubsection{Additional Knowledge}
The main output of document similarity module is the set of triples 
in form of $\langle d_i,d_j,sim_{i,j}\rangle$, 
where $d_i$ and $d_j$ are documents and $sim_{i,j}$ 
denotes the similarity between them. 
However, during the execution of this module, an additional output is generated. It contains potentially useful information such as:

\begin{itemize}
\item top N terms with the highest frequencies that might be considered as additional stop words
\item top N terms with the highest importance to a given document
\item top N articles with the lowest and highest number of distinct words.
%\item similar documents written by different authors
%\item similar articles from different categories or domains.
\end{itemize}
    
\subsection{Document Classification - General Description}
\label{sec:document_classification}
Besides a natural application of document similarity to a basic, unpersonalized recommendation system, it may also be used in a document classification based on the k-nearest neighbors algorithm. 

\begin{figure}
\begin{center}
\includegraphics[width=\linewidth]{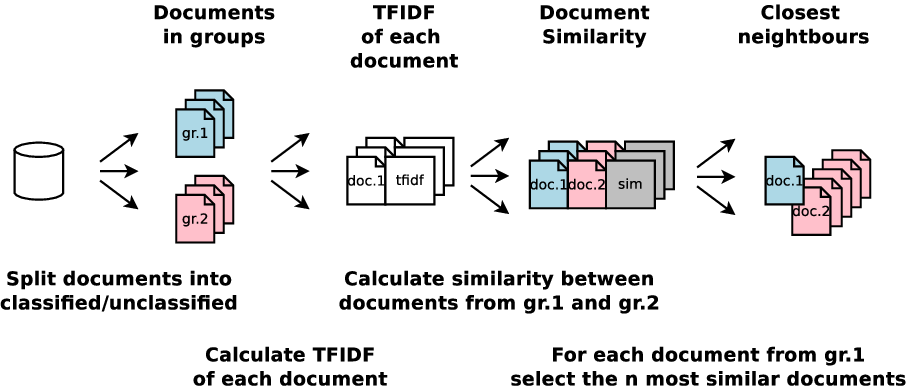}
 \end{center} 
\caption{\textbf{The initial phase for Model Creation (MC) and Classification Code Assignment (CCA).} At first documents are split into ``classified'' and ``unclassified'' group (for MC both groups contains the same metadata) and TFIDF measure is calculated over the whole set. Then, cosine similarity is calculated between documents in each group and n most similar ``classified'' documents are retained.}
\label{doc_classif_core}
\end{figure}

In the context of document classification, it is important to distinguish two topics - a model creation (MC) and a classification code assignment (CCA), each of which starts in the same way, as depicted in Fig.\ref{doc_classif_core}. In the first step, documents are split into two groups - classified and unclassified (in case of MC both groups contain the same documents). Then, TFIDF is calculated for both of these groups. Finally, document similarity between groups is calculated (excluding self-similarity) and for each document from unclassified group, \textit{n} closest neighbors are retained. After this initial phase, the subsequent step is different for MC and CCA. 

For MC, the classification codes from neighbors of a document are extracted and counted. Then, for each classification code, the best threshold is selected against given criteria, e.g. an accuracy or a precision to fit ``unclassified'' documents classification codes. Finally, the pairs $\langle$classification code, threshold$\rangle$ are persisted.

In case of CCA, after extraction of the classification code, the number of classification code occurrences are compared with a classification code threshold from a model and retained if greater or equal.

\subsection{Document Classification - Implementation Details}
\subsubsection{Sequence of Operations}
The amount of data to be transferred between Hadoop nodes has a great influence on the performance of the whole workflow. Therefore, operations depicted in Fig.\ref{doc_classif_core} should be considered in two dimensions. First one is the TFIDF calculation, for which only documents' metadata is needed. Subsequently, only information about document ID and its TFIDF are needed. The second dimension refers to the splitting into subsets ``unclassified''/``classified'' or into folds for the sake of n-fold cross validation. Because division operations can be collapsed into one, it is important to put all of them in the first place, followed by the document similarity calculations, and do not place TFIDF calculation in between.

\subsubsection{Language Choice}
For data scientists dedicated to implement and enhance algorithms in MapReduce, it is crucial to take advantage of programming languages created specifically for MapReduce. Again, Apache Pig looms as the natural candidate for document classification. Besides its strengths (predefined functions, UDFs), it should be noted that Pig (as a MapReduce paradigm) lacks some general purpose instructions like loops or conditional statements. However, it is easy to encapsulate Pig scripts into workflow management tools such as Apache Oozie or simply to use Bash shell which offers such operations. Moreover, due to the presence of macro and import statements, one can abbreviate a size of description by extracting popular transformations into macros and inserting them into separate files. In this approach, a variant of an operation (e.g. a way of calculating document similarity) can be passed to a general script as a parameter used in the import statement.

For the sake of optimization of memory utilization and calculation speed improvement it is important to use specialized types of a general operation. In case of \textit{join} operation, there are dedicated types for joining small data with a large one (``replicated join''), joining data with a mixed, undetermined size (``skewed join'') and joining sorted data (``merge join'').

\subsubsection{Data Storage}
The most utilized ways of storing data in the Apache Hadoop ecosystem are Hadoop database - HBase and Hadoop file system - HDFS. When massive data calculations are considered, then the better choice is the HDFS. When many calculating units are trying to connect to the HBase, then not all of them may be served before timeout expires. That results in a chain of failures (tasks assigned to calculation units are passed from failed to working ones, which become more and more overwhelmed by the amount of data to process). On the other hand, such failure cannot happen when the HDFS is used. 

Using HDFS in MapReduce jobs requires pre-packing data into SequenceFiles, which store data in the form of $\langle$Key,Value$\rangle$ pair. To obtain the most generic form, it is recommended to collect the key and value objects as a BytesWritable class, where a value object contains data serialized as ProtocolBuffers. This approach makes it easy to store and extend schema of any kind of data. Our experience is that reading and writing $\langle$BytesWritable,BytesWritable$\rangle$ pairs, opposed to Java and Scala usage, results in some complications in Apache Pig v.0.9.2. In that case, one may consider to encapsulate BytesWritable into NullableTuple class.

\subsubsection{Workflow Management} 
As mentioned previously, one of the best way to build a chain of data transformations is to employ a workflow manager or a general purpose language. The experiences with employing Apache Oozie and Bash were strongly in favour of the former one. Apache Oozie is a mature solution, strongly established in the Apache Hadoop ecosystem, aimed for defining and executing (when triggered by a user, time event or data arrival) workflows. In fact, using Bash or Python would require a burden of implementing Apache Oozie-like tool e.g. for the persistence of an execution history.

\section{Summary and Future Work}
\label{sec:summary}
In this article we have described the experience gained in the implementation of CoAnSys framework. Decisions we took in the development process required about half a year of tries and failures. It is hard to find coherent studies of different algorithms' implementations and therefore we hope that this contribution can save time of people and institutions preparing to embrace MapReduce paradigm and especially Apache Hadoop ecosystem into data mining systems.

This description is the snapshot of an on-going work, hence many more improvements and observations are expected to be done in a future.

%\begin{minipage}{\linewidth}
\bibliographystyle{splncs}
\bibliography{bibliography}

\begin{thebibliography}{10}

\bibitem{McKinsey}
Manyika, J., Chui, M., Brown, B., Bughin, J., Dobbs, R., Roxburgh, C., Byers,
  A.H.:
\newblock {Big data: The next frontier for innovation, competition, and
  productivity}.
\newblock Technical report, Mc Kinsey (May 2011)

\bibitem{MentalEq}
Dendek, P.J., Czeczko, A., Fedoryszak, M., Kawa, A., Wendykier, P., Bolikowski,
  L.:
\newblock How to perform research in hadoop environment not losing mental
  equilibrium - case study.
\newblock arXiv:1303.5234 [cs.SE] (2013)

\bibitem{Ferrucci2004}
Ferrucci, D., Lally, A.:
\newblock {UIMA: an architectural approach to unstructured information
  processing in the corporate research environment}.
\newblock Natural Language Engineering \textbf{10}(3-4) (2004)  327--348

\bibitem{Bembenik:1503879}
Bembenik, R., Skonieczny, L., Rybinski, H., Niezgodka, M.:
\newblock {Intelligent Tools for Building a Scientific Information Platform}.
\newblock Studies in Computational Intelligence. Springer, Berlin (2012)

\bibitem{Manghi2010}
{Manghi P.}, M.N.H.W., Peters, D.:
\newblock {An infrastructure for managing EC funded research output - The
  OpenAIRE Project}.
\newblock The Grey Journal (TGJ) : An International Journal on Grey Literature
  \textbf{6} (2010)  31--40

\bibitem{Manghi2012}
Manghi, P., Bolikowski, L., Manola, N., Schirrwagen, J., Smith, T.:
\newblock Openaireplus: the european scholarly communication data
  infrastructure.
\newblock D-Lib Magazine \textbf{18}(9/10) (2012)

\bibitem{DendekBL:2012:IAPR}
Dendek, P.J., Bolikowski, L., Lukasik, M.:
\newblock {Evaluation of Features for Author Name Disambiguation Using Linear
  Support Vector Machines}.
\newblock In: Proceedings of the 10th IAPR International Workshop on Document
  Analysis Systems. (2012)  440--444

\bibitem{Dendek2013}
Dendek, P.J., Wojewodzki, M., Bolikowski, L.:
\newblock Author disambiguation in the yadda2 software platform.
\newblock In Bembenik, R., Skonieczny, L., Rybinski, H., Kryszkiewicz, M.,
  Niezgodka, M., eds.: Intelligent Tools for Building a Scientific Information
  Platform. Volume 467 of Studies in Computational Intelligence.
\newblock Springer Berlin Heidelberg (2013)  131--143

\bibitem{BolikowskiD:2011}
Bolikowski, L., Dendek, P.J.:
\newblock {Towards a Flexible Author Name Disambiguation Framework}.
\newblock In Sojka, P., Bouche, T., eds.: Towards a Digital Mathematics
  Library, Masaryk University Press (2011)  27--37

\bibitem{Tkaczyk2012}
Tkaczyk, D., Bolikowski, L., Czeczko, A., Rusek, K.:
\newblock A modular metadata extraction system for born-digital articles.
\newblock In: Document Analysis Systems (DAS), 2012 10th IAPR International
  Workshop on. (march 2012)  11--16

\bibitem{Lukasik2013}
Lukasik, M., Kusmierczyk, T., Bolikowski, L., Nguyen, H.:
\newblock Hierarchical, multi-label classification of scholarly publications:
  Modifications of ml-knn algorithm.
\newblock In Bembenik, R., Skonieczny, L., Rybinski, H., Kryszkiewicz, M.,
  Niezgodka, M., eds.: Intelligent Tools for Building a Scientific Information
  Platform. Volume 467 of Studies in Computational Intelligence.
\newblock Springer Berlin Heidelberg (2013)  343--363

\bibitem{Kusmierczyk2012}
Kusmierczyk, T.:
\newblock Reconstruction of msc classification tree.
\newblock Master's thesis, The University of Warsaw (2012)

\bibitem{Fedoryszak2013}
Fedoryszak, M., Bolikowski, L., Tkaczyk, D., Wojciechowski, K.:
\newblock Methodology for evaluating citation parsing and matching.
\newblock In Bembenik, R., Skonieczny, L., Rybinski, H., Kryszkiewicz, M.,
  Niezgodka, M., eds.: Intelligent Tools for Building a Scientific Information
  Platform. Volume 467 of Studies in Computational Intelligence.
\newblock Springer Berlin Heidelberg (2013)  145--154

\bibitem{Matfed:2013}
Fedoryszak, M., Tkaczyk, D., Bolikowski, L.:
\newblock Large scale citation matching using apache hadoop.
\newblock arXiv:1303.6906 [cs.IR] (2013)

\bibitem{Lin2012}
Lin, J.:
\newblock {MapReduce is Good Enough? If All You Have is a Hammer, Throw Away
  Everything That's Not a Nail!}
\newblock (September 2012)

\bibitem{ElmagarmidIV:2007}
Elmagarmid, A., Ipeirotis, P., Verykios, V.:
\newblock {Duplicate Record Detection: A Survey}.
\newblock IEEE Transactions on Knowledge and Data Engineering \textbf{19}(1)
  (January 2007)  1--16

\bibitem{Manning2008}
Manning, C.D., Raghavan, P., Schtze, H.:
\newblock Introduction to Information Retrieval.
\newblock Cambridge University Press, New York, NY, USA (2008)

\bibitem{MRAlgsCloudera}
Cloudera:
\newblock Mapreduce algorithms.
\newblock
  \url{http://blog.cloudera.com/wp-content/uploads/2010/01/5-MapReduceAlgorith%
ms.pdf} (2009)

\bibitem{LeeHerKim:2011}
Lee, H., Her, J., Kim, S.R.:
\newblock Implementation of a large-scalable social data analysis system based
  on mapreduce.
\newblock In: Computers, Networks, Systems and Industrial Engineering (CNSI),
  2011 First ACIS/JNU International Conference on. (May)  228--233

\bibitem{WanYuXu:2009}
J.~Wan, W.~Yu, X.X.:
\newblock Design and implement of distributed document clustering based on
  mapreduce.
\newblock In: Proceedings of the Second Symposium International Computer
  Science and Computational Technology (ISCSCT). (2009)  278--280

\bibitem{Porter:1997}
Porter, M.F.:
\newblock Readings in information retrieval.
\newblock Morgan Kaufmann Publishers Inc., San Francisco, CA, USA (1997)
  313--316

\bibitem{Elsayed:2008:PDS:1557690.1557767}
Elsayed, T., Lin, J., Oard, D.W.:
\newblock Pairwise document similarity in large collections with mapreduce.
\newblock In: Proceedings of the 46th Annual Meeting of the Association for
  Computational Linguistics on Human Language Technologies: Short Papers.
  HLT-Short '08, Stroudsburg, PA, USA, Association for Computational
  Linguistics (2008)  265--268

\end{thebibliography}
%\end{minipage}
 
\end{document}